\documentclass[oneside,british]{elsart}
\usepackage[T1]{fontenc}
\usepackage[latin9]{inputenc}
\usepackage[a4paper]{geometry}
\usepackage{textcomp}
\usepackage{amstext}
\usepackage{graphicx}
\usepackage{amssymb}
\usepackage{esint}
\usepackage[authoryear]{natbib}

\makeatletter

\newcommand{\lyxmathsym}[1]{\ifmmode\begingroup\def\b@ld{bold}
  \text{\ifx\math@version\b@ld\bfseries\fi#1}\endgroup\else#1\fi}

\usepackage{hyperref}

\@ifundefined{showcaptionsetup}{}{%
 \PassOptionsToPackage{caption=false}{subfig}}
\usepackage{subfig}
\makeatother

\usepackage{babel}

\begin{document}
\begin{frontmatter}

\title{Sublimation of the Martian $\mathrm{CO_{2}}$ Seasonal South Polar
Cap}

\author{Schmidt Frederic$^{1,2,\boxtimes}$, Schmitt Bernard$^{3}$, Doute
Sylvain$^{3}$, Forget Francois$^{4}$,}

\author{Jian Jeng-Jong$^{5}$, Martin Patrick$^{1}$, Langevin Yves$^{6}$,
Bibring Jean-Pierre$^{6}$,}

\author{ and  the OMEGA Team}

\address{$^{1}$ ESAC, PO Box 78, E-28691 VILLANUEVA DE LA CANADA, MADRID,
SPAIN}

\address{$^{2}$ at present time: IDES, UMR 8148, Universite Paris Sud, CNRS,
Batiment 509 Universite Paris-Sud 91405 Orsay, France}

\address{$^{\boxtimes}$ Corresponding author, E-mail: frederic.schmidt@u-psud.fr,
Phone: (+33) 1 69 15 61 52, Fax: (+33) 1 69 16 48 82}

\address{$^{3}$ Laboratoire de Planetologie de Grenoble, Universite de Grenoble,
CNRS, INSU, BP 53, 38041 Grenoble Cedex 9, France}

\address{$^{4}$ Laboratoire de Meteorologie Dynamique, Institut Pierre Simon
Laplace, Universite Paris VI, CNRS, INSU, BP99, 75252 Paris Cedex
05, France}

\address{$^{5}$ Institute of Astronomy, National Central University, 10F,
No. 439, Dunhua Road, Beitun District, Taichung City 406, Taiwan}

\address{$^{6}$ Institut d'Astrophysique Spatiale, Universite Paris XI, CNRS,
INSU, Batiment 121, 91405 Orsay cedex, France}


\begin{abstract}
The polar condensation/sublimation of $\mathrm{CO_{2}}$, that involve
about one fourth of the atmosphere mass, is the major Martian climatic
cycle. Early observations in visible and thermal infrared have shown
that the sublimation of the Seasonal South Polar Cap (SSPC) is not
symmetric around the geographic South Pole. 

Here we use observations by OMEGA/Mars Express in the near-infrared
to detect unambiguously the presence of $\mathrm{CO_{2}}$ at the
surface, and to estimate albedo. Second, we estimate the sublimation
of $\mathrm{CO_{2}}$ released in the atmosphere and show that there
is a two-step process. From Ls=180\textdegree{} to 220\textdegree{},
the sublimation is nearly symmetric with a slight advantage for the
cryptic region. After Ls=220\textdegree{} the anti-cryptic region
sublimation is stronger. Those two phases are not balanced such that
there is $22\%\pm9$ more mass the anti-cryptic region, arguing for
more snow precipitation. We compare those results with the MOLA height
measurements. Finally we discuss implications for the Martian atmosphere
about general circulation and gas tracers, e.g. Ar.
\end{abstract}
\end{frontmatter}

\section{Introduction}

\begin{figure}
\center\includegraphics[bb=0bp 0bp 612bp 612bp,clip,width=10cm]{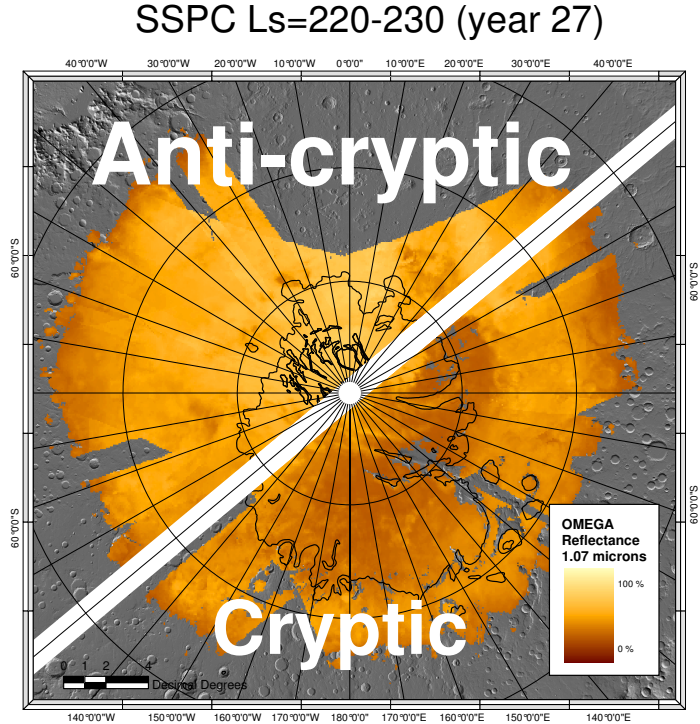}

\caption{Definition of the cryptic and anti-cryptic region within this article.}

\label{fig:DefCrypticAnti-cryptic}
\end{figure}

The \textquotedblleft{}cryptic region\textquotedblright{} is a dark
region covered by ice in the South Polar Region of Mars (\citet{Kieffer_SouthRecessionTes_JGR2000,Titus_MarsPolarProcess_Book2008}).
It appears to be a region where most of the {}``spiders'' features
are located (\citet{Piqueux_SublimationSpiderCryptic_JGR2003,Kieffer_ClodJets_JGR2007}).
In order to simplify, we define in this article the cryptic region
from longitude 50\textdegree{}E to 230\textdegree{}E (through longitude
90\textdegree{}E), and the anti-cryptic region from longitude 130\textdegree{}W
to 50\textdegree{}E (the complementary sector, passing through longitude
0\textdegree{}E) (see fig. \ref{fig:DefCrypticAnti-cryptic}).

During the phase of $\mathrm{CO_{2}}$ frost accumulation, the Seasonal
South Polar Cap (SSPC) is mainly formed by direct condensation, but
some snow events can occur (\citealp{Forget_CO2snow_Icarus1998}).
GCM studies show that a topographic forcing by the Hellas basin creates
an asymmetry in the mode of deposition. Precipitation events are more
frequent for the anti-cryptic sector than for the cryptic sector (\citealp{Colaprete_AlbedoSouthPole_Nature_2005,Giuranna_PfsCondensingSSPC_Icarus2008}).
This suggests that the texture should be more granular (smaller grain
sizes) for the former compared to the latter. This texture difference
produces a relatively higher albedo in the anti-cryptic region and
also possibly a higher accumulated mass. 

The direct measurement of the sublimating $\mathrm{CO_{2}}$ mass
on Mars has been done only recently by three different techniques:
gravity (\citet{Smith_SaisonalSnowDepth_Science_2001,Karatekin_globalCO2_JGR2006}),
neutron flux (\citet{Litvak_PolarRegionsHEND-Odyssey_Icarus_2006})
and gamma ray flux (\citet{Kelly_CO2-GRS_JGR2006}). But the time
and space resolutions of these methods prohibit any conclusion at
the regional scale. At the present time, only direct measurements
of the seasonal variations in altitude by MOLA are able to estimate
the local $\mathrm{CO_{2}}$ mass sublimation (\citet{Smith_SaisonalSnowDepth_Science_2001,Aharonson_Co2sheetMOLA_jgr2004,Jian_MOLAcryptic_ASR2009}).
But this impressive technique is limited by low signal to noise ratio
because seasonal topographic variations are typically less that one
meter! To enhance the climatic signal, \citet{Aharonson_Co2sheetMOLA_jgr2004}
use a Fourier transform and apply a filter on the MOLA time variation
keeping only the annual period. Results show that the amplitude of
this annual Fourier cycle is asymmetric, with a maximum in the anti-cryptic
region (see fig. 4 (a) in \citet{Aharonson_Co2sheetMOLA_jgr2004}).
This results lead to two different conclusions: (i) asymmetry in amplitude
is actually related to asymmetry in accumulation and/or (ii) the annual
period filtering may be not relevant to describe the real annual mass
variation, so that the amplitude asymmetry may be an artefact related
to the time of presence of $\mathrm{CO_{2}}$. The second hypothesis
is supported by the fact that the sublimation is much faster in the
cryptic region (\citet{Kieffer_SouthRecessionTes_JGR2000,Langevin_SSPC_JGR2007})
leading to lower amplitude of the annual Fourier cycle. If this hypothesis
is valid, the bi-annual cycle should be stronger for the cryptic region.
More recently, \citet{Jian_MOLAcryptic_ASR2009} focus on the south
polar region and show that cryptic and non-cryptic regions did not
present any significant thickness difference.

Indirect measurement of the $\mathrm{CO_{2}}$ mass has been performed
for decades through modeling of: (i) the recession of the surface
covered by ice observed by visible/infrared detectors (\citet{Leighton1966,Paige1985print,1993JGR....98.3339L,Forget1998print,Kieffer_SouthRecessionTes_JGR2000,Schmidt_AlbedoSSPC_Icarus2009}),
(ii) fitting the annual surface pressure cycle (\citet{PollackJGR1990,Wood1992,Hourdin_GCMpressureViking_JGR1995}).
The first modeling methodology has the advantage to allow simple estimation
of $\mathrm{CO_{2}}$ ice sublimation mass at regional scale. It assumes
that the $\mathrm{CO_{2}}$ ice is in vapour-pressure equilibrium
with the local atmosphere, and that the $\mathrm{CO_{2}}$ mass balance
at the surface is controlled by radiative balance. We propose to use
this methodology using the latest OMEGA observation to estimate the
local $\mathrm{CO_{2}}$ mass balance in order to derive the difference
in the cryptic/anti-cryptic region.

\begin{figure}
\center\includegraphics[bb=0bp 120bp 792bp 612bp,clip,width=10cm]{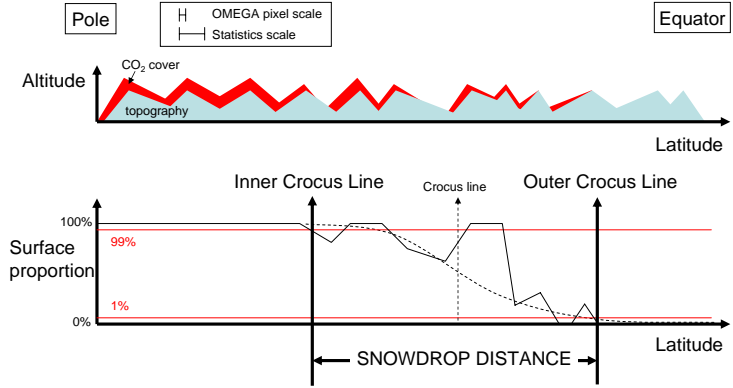}

\caption{Definition of inner crocus line, outer crocus line and snowdrop distance.
Statistics scale length $L_{stat}$ define the super pixel resolution,
binning several OMEGA pixels. Black line represents the actual surface
proportion covered by $\mathrm{CO_{2}}$ ice for each super pixel.
Doted lines represents the arctan model law (\citet{Kieffer_SouthRecessionTes_JGR2000}).
Crocus line is defined in the inflection point of the arctan law.
Inner and Outer crocus lines are defined by the surface proportion
of 99\% and 1\%.}

\label{fig:SchemaCrocusSnowdrop}
\end{figure}

For each considered location on the map, the so-called crocus date
is the date (\citet{Kieffer_SouthRecessionTes_JGR2000}) when the
surface deposit is completely gone. The crocus date is then a geophysical
field of importance for atmospheric studies. Due to the fact that
all instruments have a limited spatial resolution, \citet{Schmidt_AlbedoSSPC_Icarus2009}
has introduced both inner and outer crocus date. Each location is
defined by a super pixel with a spatial extent defined by the $L_{stat}$
scale length (with a the pixel scale length $L_{pix}$<$L_{stat}$).
The inner crocus date - respectively outer crocus date - is the date
when the first - resp. the last - pixels  fails to exhibit any ice
signature within the considered super pixel (see fig. \ref{fig:SchemaCrocusSnowdrop}). 

We will use the D-frost model (\citet{Schmidt_AlbedoSSPC_Icarus2009,Schmidt_MOMAConf2008})
to estimates the $\mathrm{CO_{2}}$ frost mass balance. Several inputs
are required: inner and outer crocus dates, albedo, altitude, slope.
This model is able to calculate the total mass sublimated, from the
date it becomes unstable (determined by the model itself), to the
crocus date (determined by the observation). Thus the initial mass,
before net sublimation, is not required as an input but will be determined
by the model.

The aim of this work is to:
\begin{itemize}
\item estimate the main input fields, required for the D-frost model, from
OMEGA observations: albedo, inner crocus/outer crocus date.
\item calculate the sublimated mass of the SSPC using D-frost.
\item derive the height variations, assuming a constant density, and compare
to MOLA measurements.
\item estimate the spatial asymmetry of the $\mathrm{CO_{2}}$ mass release.
\item derive potential effects on circulation at global scale, and on atmospheric
gas tracers.
\end{itemize}
To avoid confusion about the sign of the {}``sublimation rate''
of the SSPC, we adopt two natural conventions: surface convention
with negative mass rates (sublimation of seasonal frosts); atmosphere
convention, with positive mass rates (pump up of the atmosphere).

\section{Methods \label{sec:Methods}}

We propose to use the OMEGA imaging spectrometer (\citet{Bibring_OMEGA_ESA_SP-v1_2004})
on board Mars Express for simultaneously detect $\mathrm{CO_{2}}$
ice on Mars and estimate the albedo under the aerosols layer. A compilation
of all dataset available is out of the scope of this article but should
be an interesting step for the future.

We use the following notation: latitude $\lambda$ , longitude $\theta$
and time $t$. The spatial fields (inner and outer crocus dates) are
sampled in a spatial grid of 0.3\textdegree{} latitude (200 steps
from $\lambda=-90\lyxmathsym{\textdegree}$ to -30\textdegree{}) and
10\textdegree{} longitude (36 steps from $\theta=0\lyxmathsym{\textdegree}$
to 360\textdegree{}). The spatio-temporal fields (albedo and sublimated
mass) are sampled in the same spatial grid and the temporal grid of
5\textdegree{}Ls (72 steps from $t=0\lyxmathsym{\textdegree}$ to
360\textdegree{}). Each element of the grid is called {}``bin''.
The succession of bins in latitude at constant longitude is called
a {}``longitude sector''. Respectively a succession of bins in longitude
at constant latitude is called a {}``latitude sector''

Due to the lack of the spatial/temporal cover we need to interpolate
the measured crocus dates and albedo. The first two sections present
our strategy. Each of these interpolation methods will be evaluated
simply using the standard deviation of the difference between the
actual data and the interpolated field.

\subsection{Crocus dates field}

For each bin with available OMEGA observation, we estimate the inner
and outer crocus dates using previously described method in \citet{Schmidt_AlbedoSSPC_Icarus2009}.
This method uses WAVANGLET, an automatic detection algorithm (\citet{Schmidt_Wavanglet_IEEETGRS2007}),
and a special algorithm to extract crocus lines. We use the OMEGA
dataset in year 2005-2006 (Martian year 27) from Ls = 110\textdegree{}
to Ls=320\textdegree{} and selected 544 observations from latitude
30\textdegree{}S to 90\textdegree{}S.

We propose here the following methodology to get an interpolated map
of both inner $t_{in}(\lambda,\theta)$ and outer $t_{out}(\lambda,\theta)$
crocus date.

\subsubsection{Interpolation method\label{sub:Interpolation-method}}

The crocus dates,  binned into our grid (see sec. \ref{sec:Methods}),
are  re-estimated by a weighted sum of actually measured neighbouring
values in order to interpolate the field in case of missing values.
The weights depend on the distance and are chosen to be in a Gaussian
shape. This operation is performed through a convolution product on
the complete latitude/longitude grid with the Gaussian kernel:\begin{equation}
t_{in}^{interp}(\lambda,\theta)=\iintop t_{in}(\lambda,\theta).\frac{1}{2\pi.\sigma_{lat}.\sigma_{long}}.\exp\left(\frac{(\lambda-\lambda')^{2}}{2\sigma_{lat}}+\frac{(\theta-\theta')^{2}}{2\sigma_{long}}\right)d\lambda'.d\theta'\label{eq:InterpolationCrocus}\end{equation}
With :
\begin{itemize}
\item  computation on the complete grid : latitude $\times$ longitude bins:
$200\times36$
\item variance of the Gaussian in latitude $\times$ longitude bins: $\sigma_{lat}=41.66$
latitude bin (12.5\textdegree{} latitude), $\sigma_{long}=5$ longitude
bin (50\textdegree{} latitude). The ratio of latitude and longitude
variance is set to real average distance in our grid.
\end{itemize}

\subsubsection{Results and evaluation}

The interpolated crocus line field $t_{in}^{interp}$, average for
the cryptic and anti-cryptic regions are drawn in fig. \ref{fig:InterpolationCrocus-Albedo}
(a). The recession is clearly symmetric before Ls=140\textdegree{},
latitude 55\textdegree{}S. This symmetry can be biased by the interpolation,
especially due to Hellas and Argyre basins where persistent $\mathrm{CO_{2}}$
ice have been detected by \citet{Giuranna_PFS-SSPC_PSS2007}. After
Ls = 190\textdegree{}, latitude 62\textdegree{}S, the cryptic region
crocus lines are 20\textdegree{}Ls earlier than the anti-cryptic region.
Two possible interpretations will be discussed in section \ref{sec:Results}:
lower accumulated $\mathrm{CO_{2}}$ ice mass and/or lower albedo
leading to higher absorption and sublimation. 

To evaluate the quality of this interpolation, we analyze the difference
between actual measurement and interpolated field. The standard deviation
is quite large (14\textdegree{}Ls for outer crocus date and 17\textdegree{}Ls
for inner crocus date) due to the sparseness and discontinuity of
the dataset. Also it could be due to the sensitivity of daily variation
in frost cover, ignored within our scheme. The mean snowdrop time
is 15\textdegree{}Ls in agreement with the value measured by consecutive
observations by \citet{Schmidt_AlbedoSSPC_Icarus2009}.

\subsection{Albedo field}

Radiative balance requires the directional-hemispheric bolometric
albedo of the surface $A$, or simply {}``albedo'', that is an integrated
quantity of the BRDF (Bidirectional Reflectance Density Function).
It is dependent of the incidence direction (the solar direction) but
integrated on all emergence directions of the top hemisphere and spectrally
integrated - weighted by the solar spectrum. Definitions can be found
in \citealp{Hapke_Book1993}. 

This quantity can be estimated from space using the reflectance spectra
in the visible/near infrared recorded by OMEGA, by removing the dust
aerosols contribution that significantly modifies the apparent surface
albedo, notably for bright icy surfaces that appear darker. \citet{Vincendon_DustAerosolsSSPC_Icarus2008}
propose a method to evaluate the contribution of the aerosol scattering
and attenuation at all wavelengths from a narrow spectral region where
the entire signal comes from the aerosols (saturated bands of $\mathrm{CO_{2}}$
ice).

We use 265 images of the Martian year 27 from Ls=120\textdegree{}
to Ls=300\textdegree{} in the latitude range 52\textdegree{}S to 90\textdegree{}S.
We average all estimations points where the method provided by \citet{Vincendon_DustAerosolsSSPC_Icarus2008}
can be applied to the corresponding bin. Then we perform an empirical
spectral integration of the solar spectrum in the visible and near-IR
as described in \citet{Schmidt_AlbedoSSPC_Icarus2009}. Finally, we
estimate the complete field of seasonal deposit albedo using the following
empirical interpolation method. The main advance of this interpolation
method compared to our previous study (\citet{Schmidt_AlbedoSSPC_Icarus2009})
is to take into account both space and time evolution.

\subsubsection{Interpolation method}

The albedo, binned into our grid (see sec. \ref{sec:Methods}), are
 re-estimated by a weighted sum of actually measured neighbouring
values in order to interpolate the field in case of missing values.
The weights, depending on the distance in space and time, are chosen
to be in a Gaussian shape. Albedo is interpolated in a similar strategy
to crocus date (section \ref{sub:Interpolation-method}) with a convolution
on the complete latitude/longitude/time grid with a Gaussian kernel.
In order to keep in a realist situation, we apply a threshold on the
resulting albedo and then smooth the results.
\begin{enumerate}
\item Convolution with a Gaussian kernel on the complete latitude/longitude/time
grid: 


\begin{eqnarray}
A^{interp1}(\lambda,\theta) & = & \iiintop A(\lambda,\theta).\frac{1}{2\pi.\sigma_{lat}.\sigma_{long}.\sigma_{time}}. \nonumber \\
& & {} \exp\left(\frac{(\lambda-\lambda')^{2}}{2\sigma_{lat}}+\frac{(\theta-\theta')^{2}}{2\sigma_{long}}+\frac{(t-t')^{2}}{2\sigma_{time}}\right)d\lambda'.d\theta'.dt'\label{eq:InterpolationAlbedo}
\end{eqnarray}

With :

\begin{itemize}
\item computation on the complete grid : latitude $\times$ longitude bins:
$200\times36\times72$
\item variance of the Gaussian in latitude $\times$ longitude $\times$
time bins: $\sigma_{lat}=41.66$ latitude bin (12.5\textdegree{} latitude),
$\sigma_{long}=5$ longitude bin (50\textdegree{} latitude), $\sigma_{time}=4$
time bin (20\textdegree{} Ls). The ratio of latitude and longitude
variance is set to the real average distance ratio in our grid.
\end{itemize}
\item Application of a threshold on the albedo $A^{interp1}$ value in the
range {[}0.3, 0.9{]} to get $A^{interp2}$ to remove noisy data. 
\item Convolution to smooth the threshold (same than eq. \ref{eq:InterpolationAlbedo})
using $A^{interp2}$ to get $A^{interp}$, with:

\begin{itemize}
\item computation on the local grid only to save computation time:  $8\times2\times2$
\item variance of the Gaussian in latitude $\times$ longitude $\times$
time bins: $\sigma_{lat}=8.3$ latitude bin (2.5\textdegree{} latitude),
$\sigma_{long}=1$ longitude bin (10\textdegree{} latitude), $\sigma_{time}=1$
time bin (5\textdegree{} Ls). The ratio of latitude and longitude
variance is set to the real average distance ratio in our grid.
\end{itemize}
\end{enumerate}

\subsubsection{Results and evaluation}

\begin{figure}
\center\subfloat[Crocus lines]{

\includegraphics[angle=90,width=8cm]{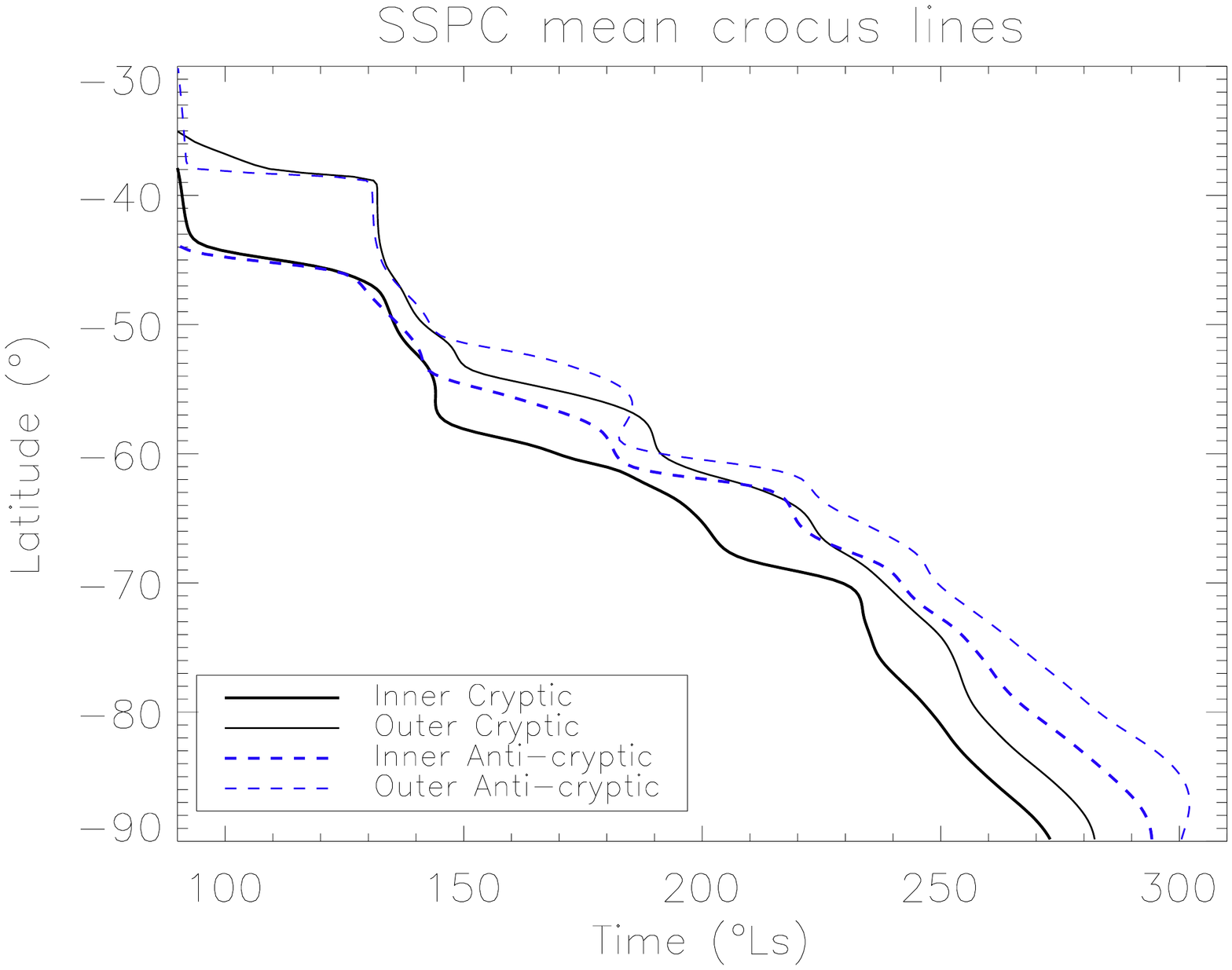}}\subfloat[Albedo]{

\includegraphics[angle=90,width=8cm]{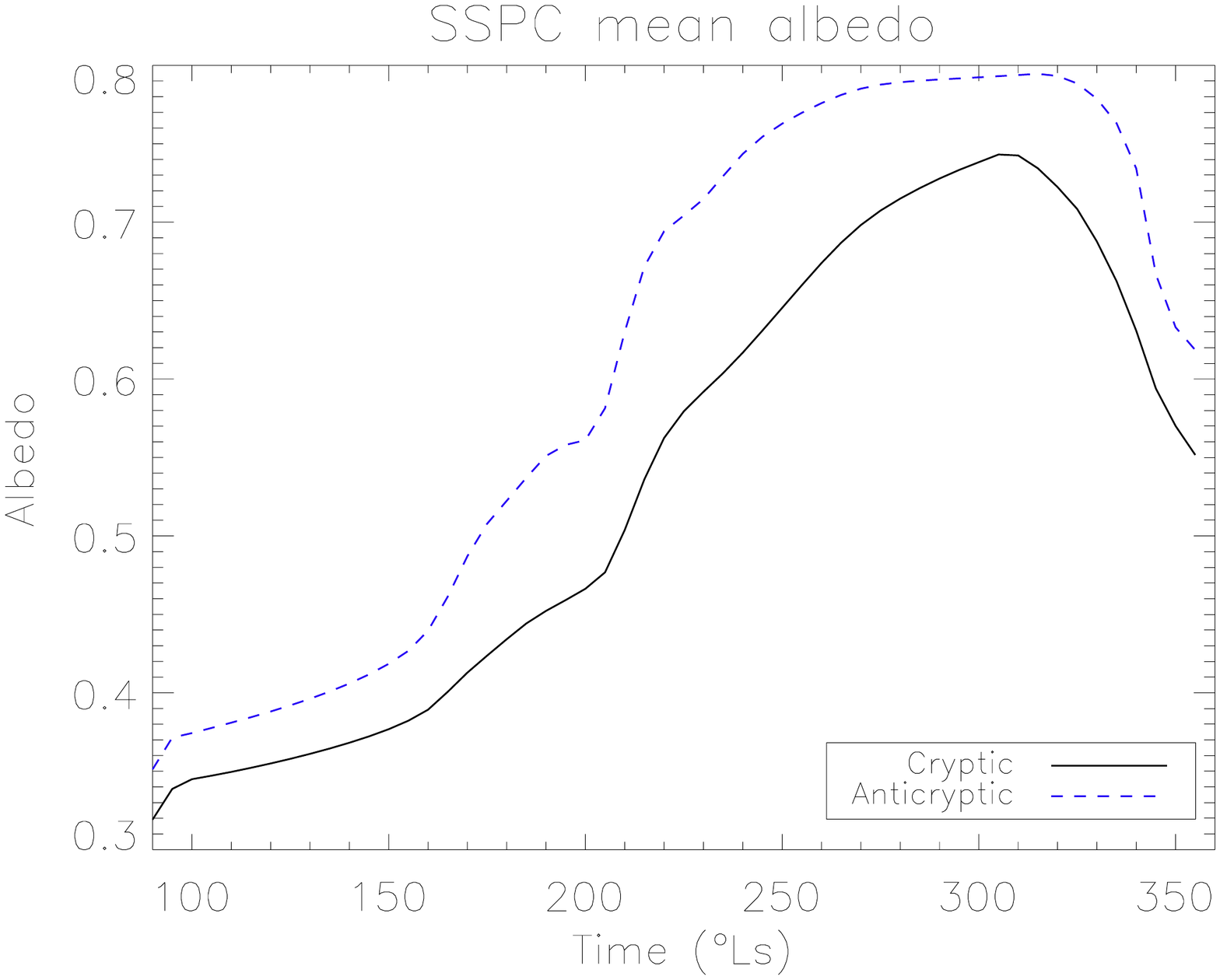}}

\caption{Averaged fields interpolated from OMEGA observations for both cryptic
and anti-cryptic regions. (a) crocus lines (b) albedo under the aerosols.
The interpolated fields are used as an input of the D-frost model.}
\label{fig:InterpolationCrocus-Albedo}
\end{figure}

The interpolated albedo field $A^{interp}$, average for the cryptic
and anti-cryptic regions are shown in fig. \ref{fig:InterpolationCrocus-Albedo}
(b). Albedo increases in both region of the SSPC as previously noted
(\citet{Paige1985print,Kieffer_SouthRecessionTes_JGR2000,Schmidt_AlbedoSSPC_Icarus2009}).
The anti-cryptic region has an albedo 0.03 higher than the cryptic
in the beginning of the recession. The difference increase to reach
0.2 around Ls=220\textdegree{} and then decrease again. The main reasons
for a lower albedo in the cryptic region, compared to the anti-cryptic
region should be the following: (i) the fact that the ice is translucent
and thus the low albedo is due to the underneath ground (\citet{Kieffer_SouthRecessionTes_JGR2000})
(ii) the presence of the fans associated with the spiders (\citet{Piqueux_SublimationSpiderCryptic_JGR2003,Kieffer_cryptic_nature2006,Kieffer_ClodJets_JGR2007})
(iii) dust contamination close to the surface (\citet{Langevin_cryptic_nature2006})
(iv) larger grain size in the case granular ice (\citet{Warren1990print}).
Recent studies suggest that a slab ice is present in extensive position
in the SSPC with a fast space and time evolution (\citet{Schmidt_SlabIce_Europlanet2009}).
We will discuss these points in an upcoming paper (\citet{Schmidt_PhysicalState_IcarusSUBMIT2009}).

This interpolation method is centred because the convolution with
a Gaussian kernel is normalized. The standard deviation is 0.028 and
is one order of magnitude lower than the mean interpolated albedo
field (0.405). The resulting empirical interpolated albedo field is
thus very close to the original measured dataset.

\subsection{D-frost model: daily average sublimated mass model}

We use the D-frost model to estimate the daily average $\mathrm{CO_{2}}$
mass surface balance $dM=dM(\lambda,\theta,t)$ for each bin 0.3\textdegree{}
latitude, 10\textdegree{} longitude and 5\textdegree{}Ls as described
in introduction of section \ref{sec:Methods}. This method estimate
the following radiative balance in the visible and IR range: \begin{equation}
dM=(F_{therm}^{out}-W_{sun}^{in}-W_{scat}^{in}-W_{IR}^{in}-W_{floor}^{in}-F_{cond}^{in})\frac{1}{L_{CO_{2}}}.dt\label{eq:FrostMassBalance}\end{equation}
$F_{therm}^{out}$ represents the thermal energy flux from the surface;
$F_{sun}^{in}$ is the direct solar energy absorbed by the surface;
$W_{scat}^{in}$ is the solar energy scattered by the atmosphere and
absorbed by the surface; $W_{IR}^{in}$ represents the thermal energy
emitted by the atmosphere and absorbed by the surface; $W_{floor}^{in}$
is the thermal energy emitted by the neighbouring surface facets and
absorbed by the surface; $F_{cond}^{in}$ is the heat flux from the
regolith; $L_{CO_{2}}=590\times10^{3}\:\mathrm{J.kg^{-1}}$ is the
latent heat for $\mathrm{CO_{2}}$ sublimation. 

The thermal energy flux is estimated with the frost temperature determined
by the local pressure of $\mathrm{CO_{2}}$, based on the LMD GCM
(\citet{Forget_MarsGCM_JGR1999}) - calibrated to Viking Lander pressure
curves -, scaled to MOLA elevation. The albedo in the thermal infrared
domain is assumed to be constant at 0.1 and the emissivity is set
to 0.99. Despite these parameters are actually changing in space and
time (\citet{Eluszkiewicz_MicrophysicsRadiativeIces_Icarus2003}),
a quantitative sensitivity test has shown that it should play a minor
role (see section 4.1.6. of \citet{Schmidt_AlbedoSSPC_Icarus2009}).
Also the annual heat wave is neglected to speed up the computation.
This energy source has also been demonstrated to be negligible during
the sublimation phase (see section 4.1.3. of \citet{Schmidt_AlbedoSSPC_Icarus2009}).
More details about the computation scheme can be found in \citet{Schmidt_AlbedoSSPC_Icarus2009}.

\subsubsection{Ar description}

We improve the D-frost model by including the effect of the non-condensable
gas on the $\mathrm{CO_{2}}$ partial pressure. We use the measurements
of Ar by \citet{Sprague_Ar-atmo-Mars_JGR2007} to correct for the
actual $\mathrm{CO_{2}}$ pressure using the formula: \begin{equation}
P_{CO_{2}}=P_{GCM}-P_{Ar}\end{equation}
In other words, we use $P_{CO_{2}}$ instead of $P_{GCM}$ in the
estimation of surface temperature (see eq. 13 of \citet{Schmidt_AlbedoSSPC_Icarus2009}).

We use the interpolated albedo field presented in the last section.
Altitude and slopes are estimated for each bin averaging a map in
stereographic south projection of MOLA data at a resolution of 920.8
m.

\subsubsection{Time integration \label{sub:Time-integration}}

The total mass sublimated $M$ is estimated by this integral:

\begin{equation}
M=\intop_{-30\lyxmathsym{\textdegree}}^{-90\lyxmathsym{\textdegree}}\intop_{0\lyxmathsym{\textdegree}}^{360\lyxmathsym{\textdegree}}\intop_{t_{0}}^{t_{end}}dM(\lambda,\theta,t).dS.dt\label{eq:totSUBmass}\end{equation}
where $t_{0}$ is the beginning of the sublimation (i.e. the first
time when $dM<0$), $t_{end}$ is the crocus date (i.e. the time when
all the $\mathrm{CO_{2}}$ ice disappears).

We propose two extreme cases for $t_{end}$: the inner crocus date
$t_{in}$ and outer the crocus line $t_{out}$. A most realistic case
is the decreasing of surface proportion $X=X(t)$ inside each bin
from the inner to the outer crocus date. Then eq. \ref{eq:totSUBmass}
becomes:

\begin{equation}
M=\intop_{-30\lyxmathsym{\textdegree}}^{-90\lyxmathsym{\textdegree}}\intop_{0\lyxmathsym{\textdegree}}^{360\lyxmathsym{\textdegree}}\intop_{t_{0}}^{t_{end}}dM(\lambda,\theta,t).X(t).dS.dt\end{equation}

We assume a linear decrease of $X(t)$ with time, from $X=100\%$
at $t_{in}$ to $X=0\%$ at $t_{out}$. Thus the surface proportion
can be written:

\begin{equation}
X(t)=\frac{t_{out}-t}{t_{out}-t_{in}}\end{equation}
In the following, this case will be referred to as {}``linear surface
proportion''.

\subsection{MOLA height variation}

The MOLA observations were performed from February 28, 1999 and June
30, 2001 in a time interval longer than one Martian year (MY 23 and
MY24). In this study, we used MOLA data from orbit 10012 to 20327
(from Ls = 103\textdegree{} to 360\textdegree{} in MY23 and then to
190\textdegree{} in MY24). Between Ls=344\textdegree{} in MY23 and
Ls=16\textdegree{} in MY24 there are no MOLA data acquired because
solar conjunction. This time interval permits to investigate the height
changes of the $\mathrm{CO_{2}}$ polar deposits.

To determine the thickness of the $\mathrm{CO_{2}}$ frost layer in
different regions as a function of time, it is most important to compute
the crossover residuals resulting from subtracting the altitude $h$
of a certain location at some value of Ls to the reference altitude
$h_{0}$ at the same location without frost. In this work about the
south polar region, the no frost reference altitude is chosen at Ls=300\textdegree{}.
We have limited our consideration to a latitudinal band between 50\textdegree{}S
and 85\textdegree{}S. Several small areas of about 0.1\textdegree{}-0.2\textdegree{}
in longitude and 0.05\textdegree{} in latitude and with enough number
of crossovers are then chosen. After computation of the average elevation
value of each small area, we calculate the residuals by subtracting
the average altitude $h$ of a certain location at some value of Ls
to the reference altitude $h_{0}$ at the same location.

\section{Results \label{sec:Results}}

\begin{figure}
\center\includegraphics[angle=90,width=10cm]{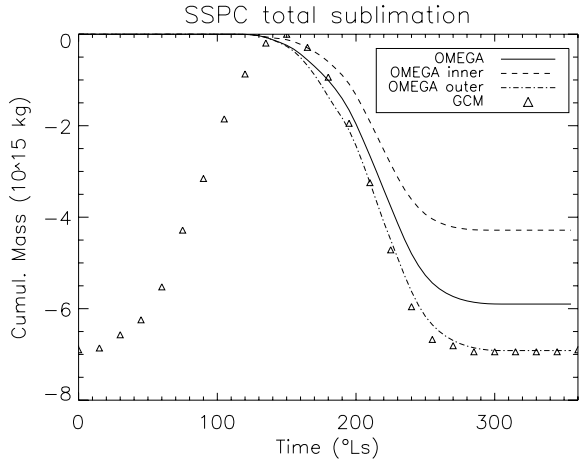}

\caption{Total cumulative sublimated mass estimated by the D-frost model using
OMEGA inputs compared with GCM (\citet{Forget_MarsGCM_JGR1999}).
Please note that a negative balance corresponds to surface mass loss
(surface convention).}
\label{fig:totalmass}
\end{figure}

The total sublimated mass from latitude 50\textdegree{}S to 90\textdegree{}S
during the year is about $5.9\times10^{15}$ kg (fig. \ref{fig:totalmass}).
The two extreme cases ($4.3\times10^{15}$ kg for $t_{end}=t_{in}$
and $6.9\times10^{15}$ for $t_{end}=t_{out}$) are considered as
the error bar due to the main uncertainty on the crocus dates such
that our estimation is $(5.6\pm1.3)\times10^{15}$ kg.

Our estimation is compatible with gravity measurements of \citet{Karatekin_globalCO2_JGR2006}
($\sim6\times10^{15}$ kg), GRS estimation from \citet{Kelly_CO2-GRS_JGR2006}
($6\times10^{15}$ kg) and HEND analysis provided by \citet{Litvak_SSPC-HEND_JGR2007}
($(6.1\pm0.6)\times10^{15}$ kg). Since our calculations are done
in the domain 50\textdegree{}S to 90\textdegree{}S only and since
we found the same numbers than other methods based on a global approach,
it suggests that the $\mathrm{CO_{2}}$ mass condensed at latitude
lower than 50S is negligible.

The GCM by \citet{Forget_MarsGCM_JGR1999}, calibrated by the pressure
variation measurement from Viking, give also an estimate of the total
SSPC of $6.9\times10^{15}$ kg, slightly higher than our estimation.
The pattern of the total sublimated mass $M$ along time (fig \ref{fig:totalmass})
is compatible with the GCM (\citet{Forget_MarsGCM_JGR1999}) although
the starting of sublimation seems to be slightly earlier for the D-frost
model. Nevertheless, there is a good agreement about the end of sublimation:
around Ls=270\textdegree{}. The shift in the beginning of sublimation
can be explained by the integration method. In this study, we integrate
spatially the $\mathrm{CO_{2}}$ mass that is subliming, independently
of the condensation domain. In the GCM integrations (\citet{Forget_MarsGCM_JGR1999}),
the whole SSPC spatial domain is covered: subliming and condensing
$\mathrm{CO_{2}}$. At Ls=90\textdegree{}, the average SSPC balance
measured by the GCM is in accumulation due to the geographic polar
region. Our integration shows already the sublimation that occurs
at low latitude. 

It is not possible to extend our calculation to the condensation domain
because one energy source is missing: heat conducted in the soil (\citet{Haberle_GroundIceCO2_PSS2008}),
which is no more negligible for the accumulation period (\citet{Schmidt_AlbedoSSPC_Icarus2009}).
Our estimation is limited in the domain latitude 50\textdegree{}S
to 90\textdegree{}S because for lower latitudes, $\mathrm{CO_{2}}$
frost may be possible only with an apparent albedo below our threshold
(0.3). Nevertheless, MOC observations shows seasonal frost at lower
latitude suspected to be $\mathrm{CO_{2}}$ ice (\citet{Schorghofer_SeasonnalFrostLowLatitude_Icarus2006}).
These condensates could be also predicted with D-frost using lower
albedo and the shadowing effect at lower scale and would be an interesting
point for the future. The good agreement between our mass balance
and direct mass measurements suggests that the low latitude frost
mass is negligible at first order.

If the Ar is well mixed in the south polar region, the effect on the
temperature and thus on the total sublimated mass is below 1\%.

\begin{figure}
\center

\subfloat[Integrated mass sublimation]{

\includegraphics[angle=0,width=10cm]{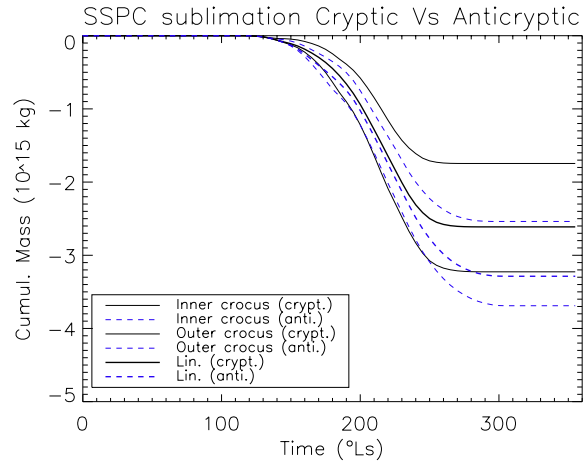}}

\subfloat[Integrated mass sublimation]{

\includegraphics[angle=90,width=10cm]{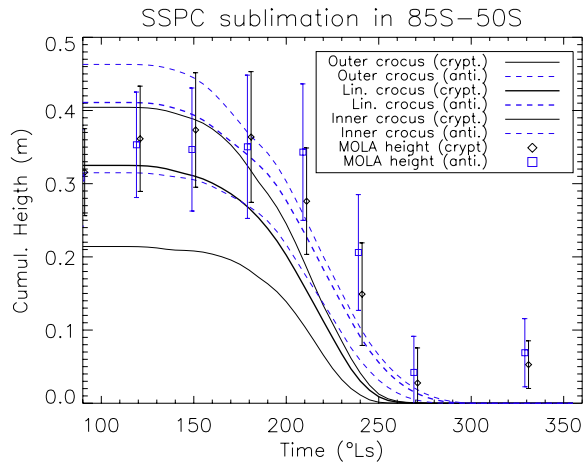}}

\caption{Total sublimated $\mathrm{CO_{2}}$ mass (a) and total sublimated
mass $\mathrm{CO_{2}}$ height (b) for both cryptic and anti-cryptic
region (surface convention) assuming a density of 920 kg.$\mathrm{m}^{-3}$
(as determined by \citet{Smith_SaisonalSnowDepth_Science_2001}) compared
to MOLA height change.}
\label{fig:mass-heigth_CryptAnti}
\end{figure}

Figure \ref{fig:mass-heigth_CryptAnti} shows the integration of the
SSPC sublimation for the two regions. The main conclusion is that
despite the asymmetry of crocus lines and albedo, the sublimated mass
seems to be symmetric around the geographic pole. The difference reach
20.5 \% (12.5 \% for $t_{end}=t_{in}$, 31.2 \% for $t_{end}=t_{out}$).
To compile all errors, we estimate roughly that the mass difference
is $22\%\pm9$. This conclusion confirms that the condensation phases
are not equivalent during the polar night for both cryptic and anti-cryptic
sectors in contrary of our previous work (\citet{Schmidt_AlbedoSSPC_Icarus2009})
argued only on four latitude points. There is a significant higher
mass condensed in the anti-cryptic region, probably due to snow event,
as previously suggested \citet{Colaprete_AlbedoSouthPole_Nature_2005,Giuranna_PfsCondensingSSPC_Icarus2008}. 

Assuming a density of 920 kg.$\mathrm{m}^{-3}$ (as determined by
\citet{Smith_SaisonalSnowDepth_Science_2001}), the estimated maximum
height from the D-frost model reach 0.32 m (0.21 m for $t_{end}=t_{in}$,
0.40 m for $t_{end}=t_{out}$) in the cryptic region and 0.41 m (0.31
m for $t_{end}=t_{in}$, 0.46 m for $t_{end}=t_{out}$) in the anti-cryptic
region (see fig. \ref{fig:mass-heigth_CryptAnti}). We compile the
errors to estimate the height $0.31\pm0.10$ m for the cryptic region
and $0.38\pm0.08$ m for the anti-cryptic region. In comparison, the
direct MOLA height measurements are $0.37\pm0.08$ m for the cryptic
region and $0.35\pm0.08$ m for the anti-cryptic region at Ls=150\textdegree{}.
At first order, absolute results from both methods are compatible
with a difference below the error bar (see fig. \ref{fig:mass-heigth_CryptAnti}).
The difference anti-cryptic/cryptic in the MOLA data is significant
at Ls=210\textdegree{} and Ls=240\textdegree{} in agreement with the
D-frost model, when the mass ratio anti-cryptic/cryptic is the highest.
At this time period, the MOLA height measurement is higher than the
D-frost estimation probably due to a $\mathrm{CO_{2}}$ bulk density
lower  than 920 kg.$\mathrm{m}^{-3}$ as suggested by \citet{Aharonson_Co2sheetMOLA_jgr2004,Giuranna_PFS-SNPC_PSS2007,Matsuo_densityCO2snow_Icarus2009}.

\begin{figure}
\center\subfloat[Absolute sublimation rates]{

\includegraphics[angle=90,width=8cm]{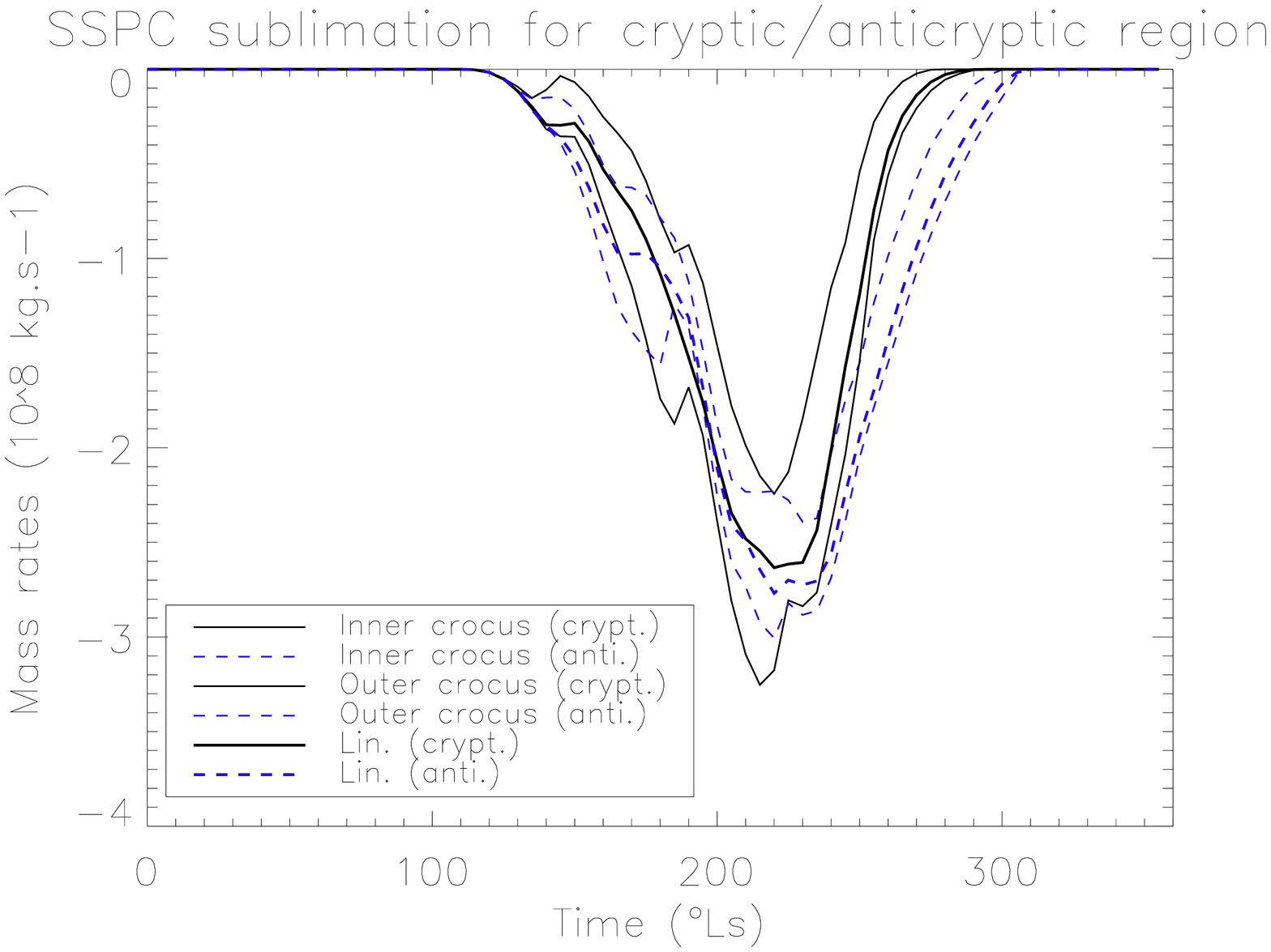}}\subfloat[Sublimation rates difference]{

\includegraphics[angle=90,width=8cm]{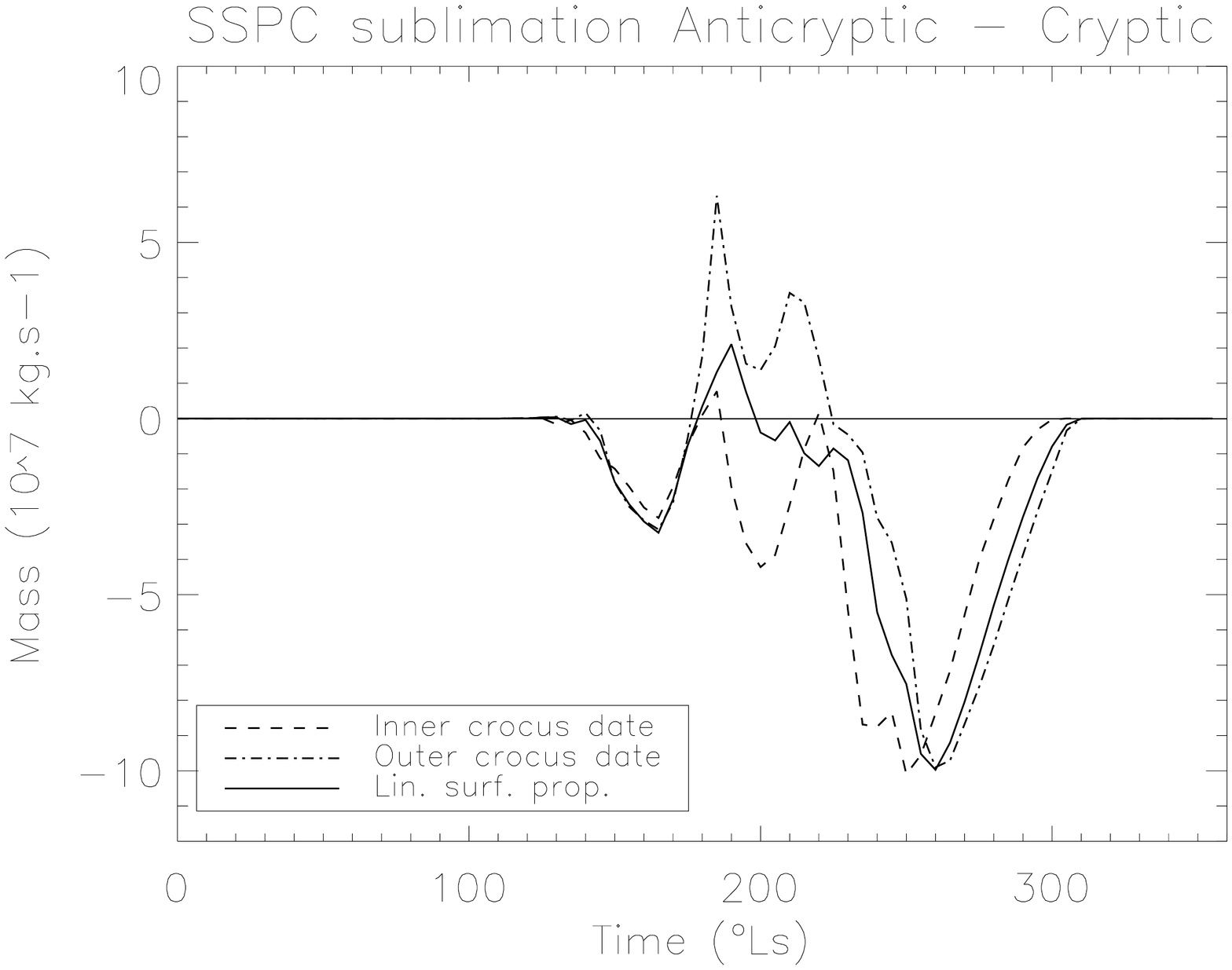}}

\caption{Sublimated Mass and within the cryptic region and the anti-cryptic
region. Note that the sublimated mass is a negative quantity so that
when this difference is positive, the cryptic region is subliming
faster than the anti-cryptic region (surface convention).}
\label{fig:DiffMassCryptAnti}
\end{figure}

During the first phase, the sublimation is nearly symmetric with a
significant higher sublimation in the cryptic region for the time
period Ls=180\textdegree{} to 220\textdegree{} (see fig. \ref{fig:DiffMassCryptAnti}(a)
and (b)). This corresponds to appearance of the low albedo cryptic
region (\citet{Kieffer_SouthRecessionTes_JGR2000,Piqueux_SublimationSpiderCryptic_JGR2003,Langevin_SSPC_JGR2007})
that absorbs and sublimates faster. 

After Ls=220\textdegree{} the average anti-cryptic sector is subliming
stronger with a maximum difference of $1.0\times10^{8}$ kg.$\mathrm{s}^{-1}$
(see fig. \ref{fig:DiffMassCryptAnti}(a) and (b)) despite a maximum
sublimation rate locally in the cryptic region. This is due to the
fact that the effect of difference in surface cover is stronger than
the effect of difference in albedo. At Ls=230, the ratio surface ratio
is $R_{surf}=\frac{Surface\: anti-cryptic}{Surface\: cryptic}=1.55$
(see fig. \ref{fig:InterpolationCrocus-Albedo} (a)) and the absorbance
ratio is $R_{abs}=\frac{1-Albedo\: anti-cryptic}{1-Albedo\: cryptic}=0.71$
(see fig. \ref{fig:InterpolationCrocus-Albedo} (b)). If we simplify
eq. \ref{eq:InterpolationAlbedo} and only keep the direct sunlight
term $W_{sun}^{in}$, the sublimation mass rates ratio can be estimated
$R_{mass}=\frac{Mass\: anti-cryptic}{Mass\: cryptic}=R_{surf}.R_{abs}=1.11$,
in agreement with our results (see fig. \ref{fig:DiffMassCryptAnti}(a)
and (b)). With this simple calculation, we demonstrate that after
Ls=220\textdegree{} the anti-cryptic region is sublimating faster
due to the fact that the SSPC recession is asymmetric, i.e.: the anti-cryptic
region reaches the outer crocus line before the cryptic region.

Two phases are not balanced such that the total mass is different
for both regions, with 20\% more mass in the anti-cryptic region,
arguing for more snow precipitation during winter time (\citet{Colaprete_AlbedoSouthPole_Nature_2005,Giuranna_PfsCondensingSSPC_Icarus2008}).
If the resulting material in the anti-cryptic region, more concentrated
in snow, has significant lower emissivity, it could also condensate
slower. If this process occurs during the polar night, 20 \% is the
minimum snow content. During winter time, {}``cold spots'' has been
observed since the 70's and been interpreted as clouds or fresh snow
at the surface (\citet{KiefferJGR1977,Forget1998print,Ivanov_CloudsMOLA_icarus2001}).
Recent studies showed that the {}``cold spots'' dynamic in thermal
infrared range is as short as 5 Julian days (\citet{Cornwall_ColdSpotsNorthMars_JGR2009,Cornwall_ColdSpot_JGR2010inpress})
suggesting that metamorphism is very active and very fast on Mars.
Thus the $\mathrm{CO_{2}}$ emissivity is not likely to be lower than
0.9 during a significant period of time.

In conclusion, our estimates indicate that the sublimation flux is
modulated by the local properties of the cryptic/anti-cryptic regions
and in particular the crocus date. \citet{Schmidt_AlbedoSSPC_Icarus2009}
conclusions are still valid: albedo is controlling the SSPC sublimation
but once there is no $\mathrm{CO_{2}}$ ice anymore it is obvious
that the sublimation stops. 

\section{Ar dilution in the south polar vortex}

Since SSPC is sublimating, the atmosphere is pumped up and rebalanced
by mass transfer at global scale (\citet{Hourdin_GCMpressureViking_JGR1995}).
This mass transfer lead also to a differential dilution of the minor
non-condensable gas that acts like tracer of the phenomena. In addition
to the sublimating SSPC, air mass coming from lower latitudes could
enter the polar vortex and dilute the local air (\citet{Colaprete_CO2cloudsCAPE_Icarus2008}).
Using a box modeling, we propose here to estimate the relevant quantity
for this system: $\mathcal{F}$ the advection flux and compare it
to $Q$ the sublimation flux.

\begin{figure}
\center\includegraphics[width=10cm]{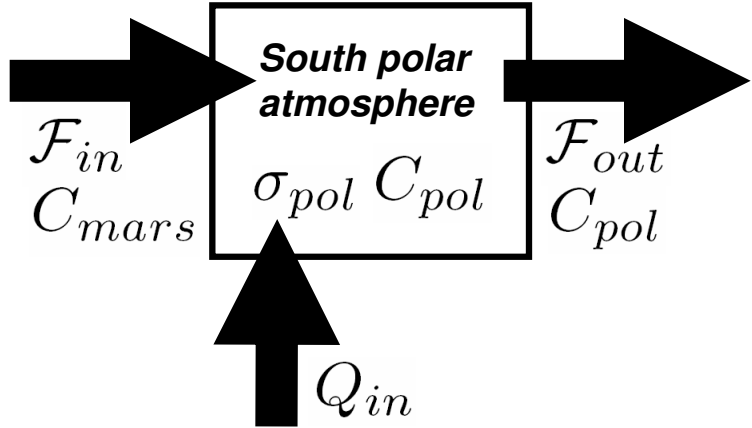}\caption{Schema of the box modeling with $\mathcal{F}$ the advection flux,
$Q$ the sublimation flux, $\sigma_{pol}$ the mass of the polar atmosphere
column and $C$ the mixing ratio of non condensable gas, such Ar.}
\label{fig:BoxSchema}
\end{figure}

Modeling the geophysical system by boxes is a very convenient approach
to understand the natural phenomena in a simple manner (\citet{Albarede_geochimie_book2009}).
We propose to model the martian atmosphere using 3 boxes: the south
polar deposits (SSPC), the polar atmosphere and the martian atmosphere
as a whole (see fig \ref{fig:BoxSchema}). 

\paragraph{$\mathrm{CO_{2}}$ balance}

We could write the mass balance of the polar atmosphere of $\mathrm{CO_{2}}$
as:\begin{equation}
\frac{d\sigma_{pol}}{dt}=Q_{in}+\mathcal{F}_{in}-\mathcal{F}_{out}\end{equation}
With $\sigma_{pol}=\frac{M_{pol}}{S_{pol}}$ the mass of the polar
atmosphere column by surface unit (in $\mathrm{kg.m^{-2}}$), the
incoming flux from the sublimation $Q_{in}=dM$ estimated in the last
section in atmosphere sign convention (in $\mathrm{kg.m^{-2}.s^{-1}}$),
$\mathcal{F}_{in}$ and $\mathcal{F}_{out}$ are respectively the
incoming and outcoming advection fluxes (in $\mathrm{kg.m^{-2}.s^{-1}}$). 

We assume that the system is in steady state $\frac{d\sigma_{pol}}{dt}=0$
such that the incoming fluxes exactly compensate the outgoing flux:
$Q_{in}+\mathcal{F}_{in}=\mathcal{F}_{out}$. In other terms, we suppose
that the mass rebalance time of the atmosphere is much lower than
the typical time of the non-condensable gas balance.

\paragraph{Non-condensable species balance}

We could also write the mass balance of the minor non condensable
species, assuming that they are in homogeneous mixing within all boxes:

\begin{equation}
\sigma_{pol}.\frac{dC_{pol}}{dt}=\mathcal{F}_{in}.C_{mars}-\mathcal{F}_{out}.C_{pol}\end{equation}
With $C_{pol}$ and $C_{mars}$, the mixing ratio of the non condensable
gas in the polar region respectively in the whole Martian atmosphere.
We solve this equation using the following numerical solution: 

\begin{equation}
C_{pol}(t+\Delta t)=C_{pol}(t).\left(1-\frac{\mathcal{F}_{in}}{\sigma_{pol}}.\Delta t-\frac{Q_{in}}{\sigma_{pol}}.\Delta t\right)+C_{mars}.\frac{\mathcal{F}_{in}}{\sigma_{pol}}.\Delta t\label{eq:conc_minor}\end{equation}
For the convenience, we use a time step $\Delta t=1s$.

Measurements of the Ar in the south polar atmosphere have been done
by the GRS instrument (\citet{Sprague_ArSouthPolar_Science2004,Sprague_Ar-atmo-Mars_JGR2007}).
During the time period of sublimation in the south polar region (Ls=110-350),
we estimate the decrease of Ar using eq. \ref{eq:conc_minor}. We
set the initial value of Ar abundance $C_{pol}(t=110\lyxmathsym{\textdegree}Ls)=3.89$\%
as measured by GRS. We use the commonly used value of the Martian
mixing ratio $C_{mars}=1.6$\% in agreement with Viking Lander 2 measurements
(\citet{1977JGR....82.4635O}).

\begin{figure}
\center\includegraphics[angle=0,width=10cm]{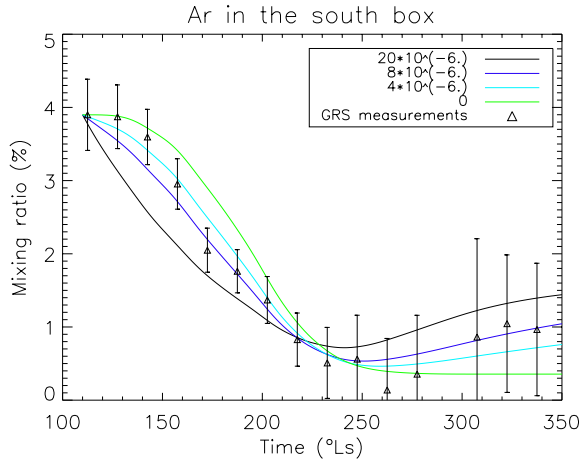}\caption{Estimation of $C_{pol}$ as a function of time in comparison with
actual Ar measurements in the South Polar atmosphere. Lines represent
the solution of our box modeling (eq. \ref{eq:conc_minor}) for four
cases: $\mathcal{F}_{in}=0$, $10^{-5}$, $5.10^{-5}$ and $10^{-4}$
$\mathrm{kg.m^{-2}.s^{-1}}$. Triangles are Ar observations of the
GRS instruments (\citet{Sprague_Ar-atmo-Mars_JGR2007}) for the whole
south polar region.}
\label{fig:abundance_mars}
\end{figure}

Minor species abundances evolution with time are plotted in fig. \ref{fig:abundance_mars},
for different values of $\mathcal{F}_{in}$: 0, $4.10^{-6}$, $8.10^{-6}$
and $20.10^{-6}$ $\mathrm{kg.m^{-2}.s^{-1}}$. The mixing ratio evolution
decreases with time to reach a minimum around Ls=250 and then increases
again. The first decrease is due to the sublimation of the seasonal
frost, free of non-condensable gas. In addition, the advection flux
dilutes Ar with a higher $\mathcal{F}_{in}$ leading to a faster dilution.
The increase in second step is due to the injection of the Martian
air, slightly enriched in Ar compared to the local polar air. The
rate of increase is higher when $\mathcal{F}_{in}$ is faster.

By matching GRS data with our box model, we estimate that the flux
$\mathcal{F}_{in}\simeq4.10^{-6}\mathrm{\: kg.m^{-2}.s^{-1}}$. This
value has the same order of magnitude as the mean value of the sublimation
flux $\left\langle Q_{in}\right\rangle =6.7\times10^{-6}\mathrm{\: kg.m^{-2}.s^{-1}}$.
This result suggests that the air mass entering the south polar vortex
during the recession phase is of the same order than the sublimated
mass. Thus the local difference in $Q_{in}$ due to asymmetry should
lead to non-negligible effect on the vortex dynamic, especially after
Ls=220\textdegree{} (see fig. \ref{fig:DiffMassCryptAnti}).

Future studies using GCM or local climate model have to be conducted
to study the actual dynamical effect of the asymmetry in sublimation.
One particular attention should be drawn on the dilution factor regarding
the dynamics in the cryptic vs anti-cryptic region. Some regional
differences should be pointed out, as suggested by GRS (\citet{Sprague_Ar-atmo-Mars_JGR2007}). 

\section{Conclusions}

We estimate the $\mathrm{CO_{2}}$ mass balance of the SSPC using
the D-frost model and estimate the total sublimated mass to be around
$(5.6\pm1,3)\times10^{15}$ kg which is compatible with gravity measurements
(\citet{Karatekin_globalCO2_JGR2006}), with measurements by gamma
ray and neutron spectroscopy (\citet{Litvak_PolarRegionsHEND-Odyssey_Icarus_2006,Kelly_CO2-GRS_JGR2006})
and with previous GCM studies (\citet{Forget_MarsGCM_JGR1999,Kelly_CO2-GRS_JGR2006}).
This agreement validates our current approach.

The effect of the Ar content, in the case of a well mixed atmosphere
is below 1\% on the total sublimated mass. A more realistic case of
denser Ar enriched air in the bottom part of the atmosphere should
be studied in the future.

We show that the SSPC is not symmetric in mass around the geographic
pole, i.e.: the cryptic and anti-cryptic regions have an accumulated
mass with a difference in order of $22\pm9$ \%. in favour of the
anti-cryptic region. Previous studies suggested that in addition to
direct condensation, during the southern fall and winter seasons,
snow precipitation is more intense in the anti-cryptic region than
in the cryptic one (\citet{Colaprete_CO2DustStorm_JGR2002,Giuranna_PfsCondensingSSPC_Icarus2008}).
Our results show that, the actual mass deposited (possibly as snow)
should be in order of 20 \%. This conclusion is still valid if the
$\mathrm{CO_{2}}$ emissivity has regional and/or temporal variations
within the range 0.9 to 1 during the spring and summer, as observed
by TES (\citet{Eluszkiewicz_MicrophysicsRadiativeIces_Icarus2003})
ans supported by fast $\mathrm{CO_{2}}$ metamorphism (\citet{Cornwall_ColdSpot_JGR2010inpress}).
The previous estimation on small regions of interests (ROI) by \citet{Kieffer_SouthRecessionTes_JGR2000}
was 852 $\mathrm{kg.m}^{-1}$ in the {}``cryptic'' ROI and 841$\mathrm{kg.m}^{-1}$
in the {}``bright cap'' ROI. The difference is probably due to the
crocus date measurement that is delayed for OMEGA in comparison to
TES. In case of subpixel mixing, due to the thermal emission dependence
on the temperature at power four, bare soil is dominating in the thermal
infrared. This effect is not present in near infrared domain and the
mixing is really linear. Further studies should be done to precisely
compare visible, near infrared and thermal infrared datasets.

From Ls=180\textdegree{} to 220\textdegree{}, we point out that the
cryptic region is sublimating slightly stronger than the anti-cryptic
region. In the second phase, anti-cryptic region sublimation is dominating
with a maximum difference in sublimation of $1.0\times10^{8}$ kg.$\mathrm{s}^{-1}$.
This results is compatible with the MOLA height measurements difference
that reach a maximum for Ls=210\textdegree{}-240\textdegree{}.

We use a simple box model to fit the GRS measurements of Ar and estimate
that the atmosphere advection flux entering the south polar vortex
is similar to the flux from the sublimating SSPC.

More precise studies on atmosphere dynamics and minor species concentration
should be carried out with a GCM or a local scale climate model. At
present time, SSPC albedo is a constant parameter in any GCM. Future
studies should be done to include the SSPC asymmetry of albedo leading
to an asymmetry of sublimation in order to simulate realistic dynamics
and dilution factors. Those results will have to be compared to the
atmospheric observations of pressure (for instance using the method
proposed by \citet{Forget_RetrievalPressureOMEGA_JGR2007,Spiga_PressureOMEGA_JGR2007})
and minor species concentration (using gamma rays, i.e.: \citet{Sprague_ArSouthPolar_Science2004},
optical measurements on the ground i.e: \citet{Encrenaz_isotopH2O-CO2_Icarus2005},
or orbiters i.e.: \citet{Melchiorri_WaterVaporOMEGA_PSS2007}).

In particular, future local climate studies should explain the actually
asymmetric Ar enrichment (\citet{Sprague_ArSouthPolar_Science2004})
that could play an indirect role in the stability of the SSPC itself.
If the atmosphere is locally enriched in Ar, the partial pressure
of $\mathrm{CO_{2}}$ is lower and then $\mathrm{CO_{2}}$ ice stability
is lower (\citet{Forget_Ar_MOMA2009}).

\section{Acknowledgments}

This work has been supported by a contract with CNES through its 'Systeme
Solaire' program. We thank the OMEGA and MOLA teams for data management.
We also thank Mathieu Vincendon providing the SSPC reflectance map,
under the aerosols. Finally, we are grateful to two anonymous reviewers
for their helpful comments. 

\bibliographystyle{elsart-harv}
\bibliography{ref}

\end{document}